\documentclass[epjST]{svjour}
\usepackage{graphics}
\begin{document}
\title{Temporal characterization of long quantum-path high harmonics.}
%\subtitle{Do you have a subtitle?\\ If so, write it here}
\author{%First author\inst{1}\fnmsep\thanks{\email{author@cnrs.fr}} \and Second author\inst{2} \and ... P.
A. Carpeggiani $^{1,3}$, B. B\'odi$^{1,4}$, E. S. Skantzakis $^{1}$, J. W. G. Tisch $^{6}$, J. P. Marangos $^{6}$,D. Charalambidis $^{1,2,5}$, P. Tzallas $^{1}$ and A. Za\"ir $^{6,7}$\fnmsep\thanks{\email{amelle.zair@kcl.ac.uk}}}
\institute{Foundation for Research and Technology, Hellas Institute of Electronic Structure and Laser, PO Box 1527, GR71110 Heraklion (Crete), Greece \and Department of Physics, University of Crete, PO Box 2208, GR71003 Heraklion (Crete), Greece\and Technische Universität Wien, Institut für PhotonikGußhausstraße 27-29, 1040 Wien, Austria \and MTA "Lend\"ulet" Ultrafast Nanooptics Group, Wigner Research Center for Physics, 1121 Budapest, Hungary  \and ELI HU Nonprofit Ltd, ELI ALPS, Dugon Ter 13, H-6720 Szeged, Hungary \and Blackett Laboratory, Imperial College London, Prince Consort Road, London SW7 2AZ, United Kingdom \and Kings College London, Department of Physics, Attosecond Physics, London WC2R 2LS United Kingdom}
\abstract{
%The temporal characterization of short and cutoff quantum paths was an important breakthrough in providing evidence that XUV high harmonic sources can reach the attosecond duration. In contrast the long quantum path temporal characterization has remained challenging due to the complex spatio-temporal characteristics of the emission arising from the long trajectories. 
We measure for the first time the duration of long-quantum path EUV high harmonics produced in xenon gas. The long-quantum path contribution to the high-harmonic signal was carefully controlled by employing a two-colour driving laser field and a high-harmonic spatial selection in the far field, over a range of $18-25\,eV$ in photon energy. To characterise the temporal profile of long quantum path high harmonics, we performed a second order volume autocorrelation ($2-IVAC$) via two EUV photon double ionization in argon. Our results show the production and characterisation of a train of EUV pulses from the long-path with pulse duration as short as $1.4\,fs$. This measurement demonstrates that the long-quantum path emission can have enough flux for performing non-linear EUV experiments, and that the long trajectories enable a pulse duration short enough to support measurements with a temporal resolution  below $2\,fs$.}
 %end of abstract
%
\maketitle
\section{Introduction}
\label{intro}
Gas phase high-order-harmonic generation (HHG) sources have been used successfully for the production of  attosecond pulse trains \cite{paul,krauzs}. These pulses are intense enough now to enable first observations of non-linear processes in matter \cite{kobayashi,papadogiannis1,papadogiannis2,miyamoto,nabekawa,benis1,benis2,midorikawa,tzallas1,manschwetus}. Observing these processes allowed XUV quantitative studies employing ion-microscopy \cite{tsatrafyllis}, XUV pulse metrology by measuring 2nd-order volume autocorrelation ($2-IVAC$) \cite{kobayashi,midorikawa,tzallas1,tzallas2,kruse,tzallas3,takahashi1}, time-resolved XUV spectroscopy studies \cite{skantzakis} and XUV-XUV pump-probe measurements of femtosecond dynamics in atoms \cite{tzallas1} and molecules \cite{carpeggiani}. 
Enhancing further the EUV-XUV intensity provided by gas phase HHG is crucial for the extension of the aforementioned studies to higher XUV photon energies and to higher order XUV driven non-linearity \cite{nayak}. The long quantum path is a good candidate to provide higher EUV-XUV pulse energies \cite{gaarde,hergott}. In the single atomic response picture, the long quantum path electronic wave packet reaches continuum (ionization time) at higher field amplitude in the optical cycle than for the short quantum path. Hence, the electronic wave packet has a higher ionization probability. In the continuum, the electronic wavepacket accumulates phase (chirp). This phase is greater for the long path than for the short path and has an opposite sign. This chirp is at the origin of the low recombination probability and the lack of temporal confinement associated with the long quantum path. However, beyond the single atomic response, phase matching conditions can be found in the interaction volume to optimize the long quantum path contribution to the HHG signal \cite{saliere,zair} as well as control its contribution to the HHG signal via laser field synthesis \cite{brugnera}. A scheme has been proposed also to compensate for the long path chirp \cite{guichard}. Therefore there is increasing interest in utilizing long path harmonics for time resolved and non-linear experiments.
As it has been shown theoretically \cite{gaarde} and experimentally \cite{kruse}, one of the signatures of the long quantum path contribution to the harmonic yield, whilst employing a single colour IR driving field, is a temporal broadening of the XUV pulses in the train. Because the duration of these long path pulses is expected to be longer than the temporal spacing of the XUV train pulses (which is equal to the half-period of the single colour driving IR laser field, e.g. 1.33 fs) the recorded 2-IVAC trace could not resolve an XUV pulse train \cite{kruse}. A way to overcome this limit, is to drive the harmonic generation process with a two-colour driving laser field. The latter configuration breaks the temporal symmetry of the interaction. One consequence of this configuration is the emission of odd and even order high harmonics every optical cycle instead of every half cycle for a single colour interaction. This configuration provides also a new degree of freedom to control the quantum paths in a such way that the XUV energy can be enhanced and the spectral phase distribution can be improved by varying the relative phase between the two colours of the driving field \cite{kim,mauritsson,brugnera,hoffmann,haessler}. 
Towards this goal, we generated high harmonics in xenon by combining a scale-invariant configuration in the loose laser focusing geometry \cite{heyl} with a two-colour synthesized driven laser field \cite{kim,brugnera} (composed of the fundamental field at 800\,nm and its orthogonally polarized second harmonic at 400\,nm). We used phase matching that optimized high-harmonic generation from the long quantum path \cite{saliere,zair} in the spectral region of 18-25\,eV and blocked the spatial on-axis emission in the far field to filter out the short path contribution.
The high flux of long path high harmonics produced is confirmed by observing two-XUV-photon-direct-double ionization (TPDDI) induced in argon gas. The TPDDI process has been used also for the temporal characterization of these pulses by performing $2-IVAC$ measurements detecting the $Ar^{2+}$ signal.
\section{Experimental scale-invariant XUV setup}
\label{sec:1}
As shown in Fig. 1, the experiment was performed utilizing a 10\,Hz repetition rate Ti:sapphire laser system, delivering 33\,fs pulses at 800\,nm carrier wavelength (IR) and energy up to 150\,mJ per pulse. The laser beam profile is shaped into an annular beam using a super-Gaussian beam-stop of 2\,cm diameter. The two-colour laser field employed to generate high-harmonics in xenon is obtained by converting $10\,mJ$  of the IR laser field into the second harmonic at 400\,nm using a $100\,\mu\,m$ BBO-typeI crystal enabling $10\,\%$ conversion efficiency. The annular two-colour beam profile, composed of the fundamental at 800\,nm and its orthogonally polarized second harmonic at 400\,nm (with energy of 1\,mJ and estimated pulse duration of 24\,fs), was focused into a xenon pulsed gas jet, for the production of odd and even high-harmonics of the 800\,nm. The intensity of the fundamental in the interaction region was set just below to the ionization saturation intensity of xenon gas ($8\cdot10^{13}\,W/cm^{2}$) at focus while the intensity of the second harmonic was estimated to be $2\cdot10^{13}\,W/cm^{2}$ at focus. In this focusing geometry the confocal parameter of the two-colour beams, $b=7 cm$, is approximately two orders of magnitude larger than the length of the xenon gas jet ($1\,mm$). Thus, the maximum intensity of the two-colour driving field along the propagation in the xenon gas can be considered as constant. The energy ratio and the relative phase between the two colours in the synthesized driving field were tuned by rotating the BBO crystal and a $400\,\mu\,m$ thick $CaF_{2}$ plate placed after the BBO crystal for fine relative delay control between the two colours. After the xenon gas jet, a silicon plate at Brewster's angle ($75^{\circ}$) was employed to  reflect the harmonics towards the detection area while attenuating the two-colour driving laser field \cite{takahashi2}. As the reflected XUV beam and the remaining part of the two-color field were spatially expanded towards the end station, a 150\,nm thick Sn filter supported by a 5\,mm aperture allowed for selecting the 11th-16th harmonics and for blocking the residual annular shaped two-color beam. The XUV beam was then focused by a split spherical gold mirror of 5\,cm focal length into an argon  gas jet which serves as detection target. 
\begin{figure}
	\centering
	\resizebox{0.85\columnwidth}{!}{%
	\includegraphics{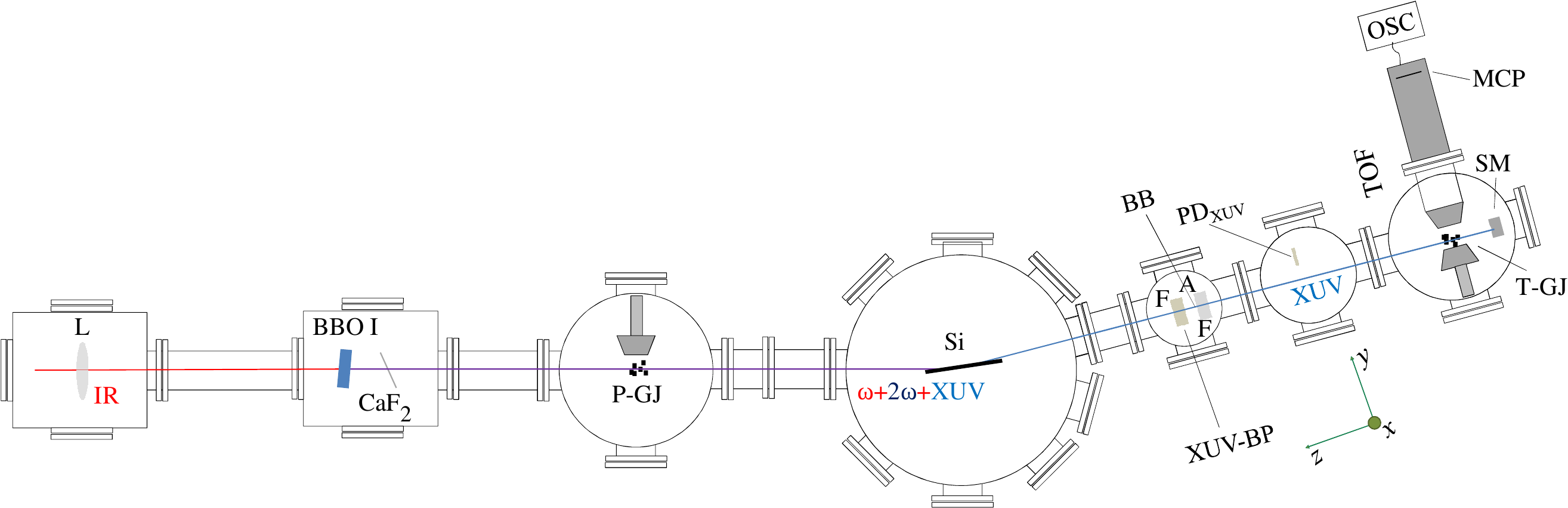} }
		\caption{Experimental scale-invariant HHG setup: (L) focusing lens of $3\,m$; synthesized two-colour field provided by a BBO crystal and a $400\,\mu\,m$ thick $CaF_{2}$; (P-GJ) pulsed gas jet; (Si) silica plate at brewster angle; (XUV-BP) XUV beam profile measurement; (BB) beam blocker of on-axis contribution; ($PD_{XUV}$) XUV photo-diode flux measurement; (F) filter;(SM) split mirror, (TOF) time of flight apparatus; (T-GJ) target TOF gas jet}
	\label{fig:Fig1}
\end{figure}
The ionization products were recorded using a $\mu$-metal shielded time-of-flight (TOF) spectrometer. The TOF can be set to record either the photoelectron energy distribution or ion-mass spectrum. The harmonic signal was maximized by adjusting the relative intensity and phase of the two-colour field which favors harmonic emission from the long path contribution ( Fig. 2a). Given the large divergence of the long quantum path far field distribution, the selection of the long path harmonics was achieved by inserting a beam block of 2.5\,mm diameter at the center of the XUV beam profile (Fig. 2b), which mainly contains the short path harmonics. Although this configuration leads to a $50\,\%$ reduction of the total XUV energy, it serves our goal to select further the long path harmonics. The energy of the long path harmonics was measured with a calibrated XUV photodiode and calculated back in the interaction region to be $\sim 8\,nJ$ per pulse in the interaction region. The intensity of the XUV radiation in the interaction region is estimated to be $\sim 5\cdot10^{11}\,W/cm^{2}$. This value has been obtained using an XUV pulse of duration $\tau_{XUV}= \tau_{L}/\sqrt{n}=15\,fs$ (where $\tau_{L}$ is the duration of the driving field and $n=3-5$ is the order of non-linearity of the harmonic generation process for harmonics in the plateau region) and assuming that the XUV focal spot diameter in the interaction region is the same as the value measured in ref.\cite{tsatrafyllis} ($4\mu\,m$). The spectrum of the long path harmonics in the interaction area was determined by measuring the energy-resolved single-photon ionization photoelectron spectra of argon gas (Fig. 2c). This is justified because in the photon energy range between 15 and 30\,eV, the single-photon ionization cross section of argon is approximately constant ($\sigma^{(1)}\approx 33\,Mb$). As the ionization potential of argon is $I_{P}=15.76\,eV$, all harmonics above the 11th contribute to the recorded single-photon photoelectron spectrum. Consequently, Fig. 2c shows the long path harmonic spectrum interacting with argon gas in the detection area. %The full understanding of all the details of structure in the harmonic spectrum would require the solution of the three-step quantum mechanical response taking into account the propagation effects of the pulsed multi-cycle two-color laser field. This is beyond the scope of the current work. 
The relative field amplitudes of the $11^{th}$ to $16^{th}$ harmonics are respectively 0.36, 0.75, 0.53, 1, 0.66 and 0.5, which correspond to a Fourier-Transform-Limited (FTL) XUV pulses in the train of 0.5\,fs.
\begin{figure}
\centering
\resizebox{0.85\columnwidth}{!}{%
	\includegraphics{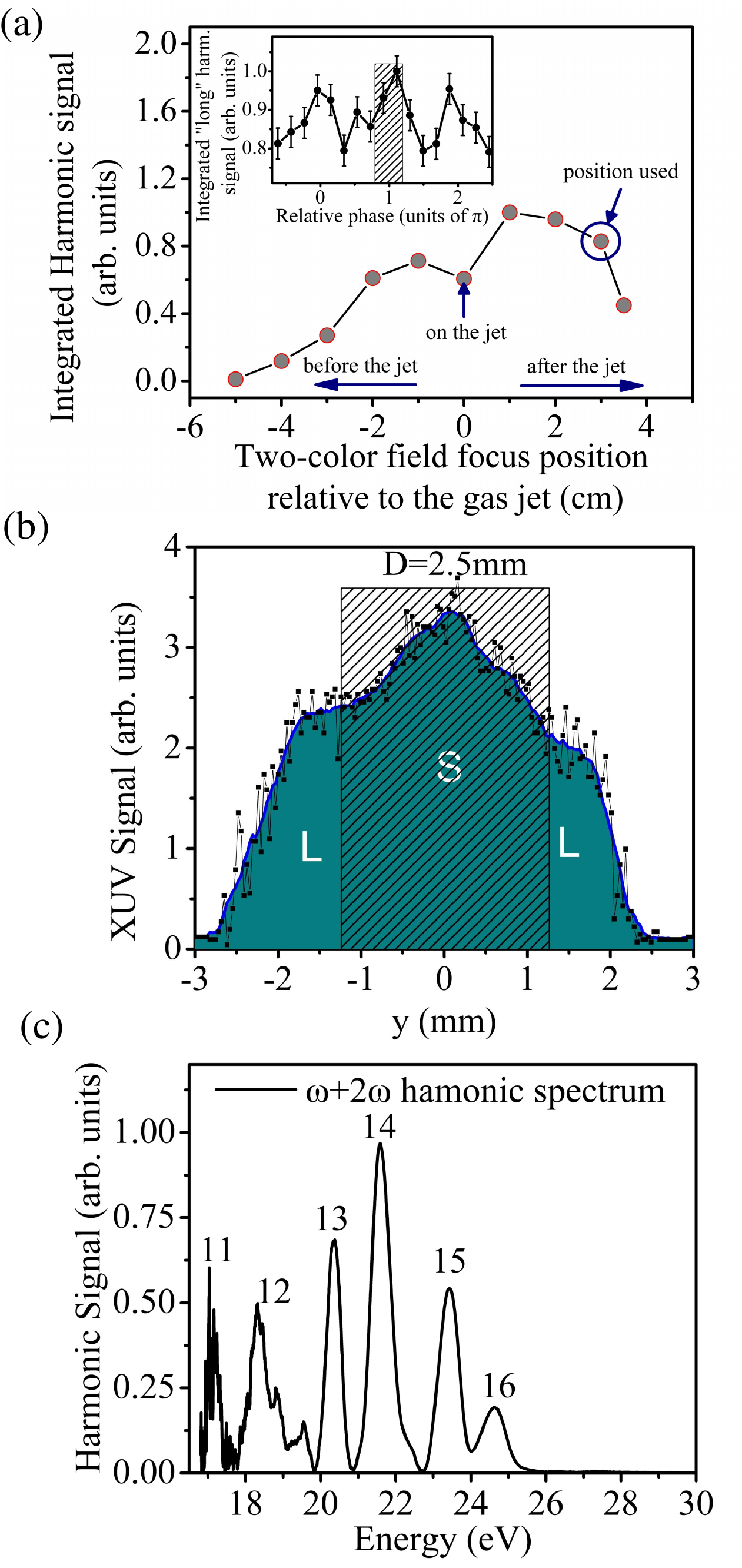} }
    	\caption{High Harmonic generation in xenon optimized for long quantum-path contribution: (a) phase matching optimum condition find by z-scan;(b) high harmonics beam profile measurement. (D) corresponds to the area which is further blocked;(c) measured odd and even high harmonic spectra from TOF}
\label{fig:Fig2}
\end{figure}
\section{Temporal characterization of the long path harmonics}
\label{sec:2}
Operating the TOF spectrometer in the ion-mass spectrometry mode, a mass spectrum which includes the $Ar^{2+}$ mass peak was recorded (Fig. 3a). The possible pathway to the production of $Ar^{2+}$ are the two-photon-direct-double ionization (TPDDI) and the three-photon-sequential double ionization (ThPSDI) (Fig. 3b).  
\begin{figure}
	\centering
	\resizebox{0.65\columnwidth}{!}{%
	\includegraphics{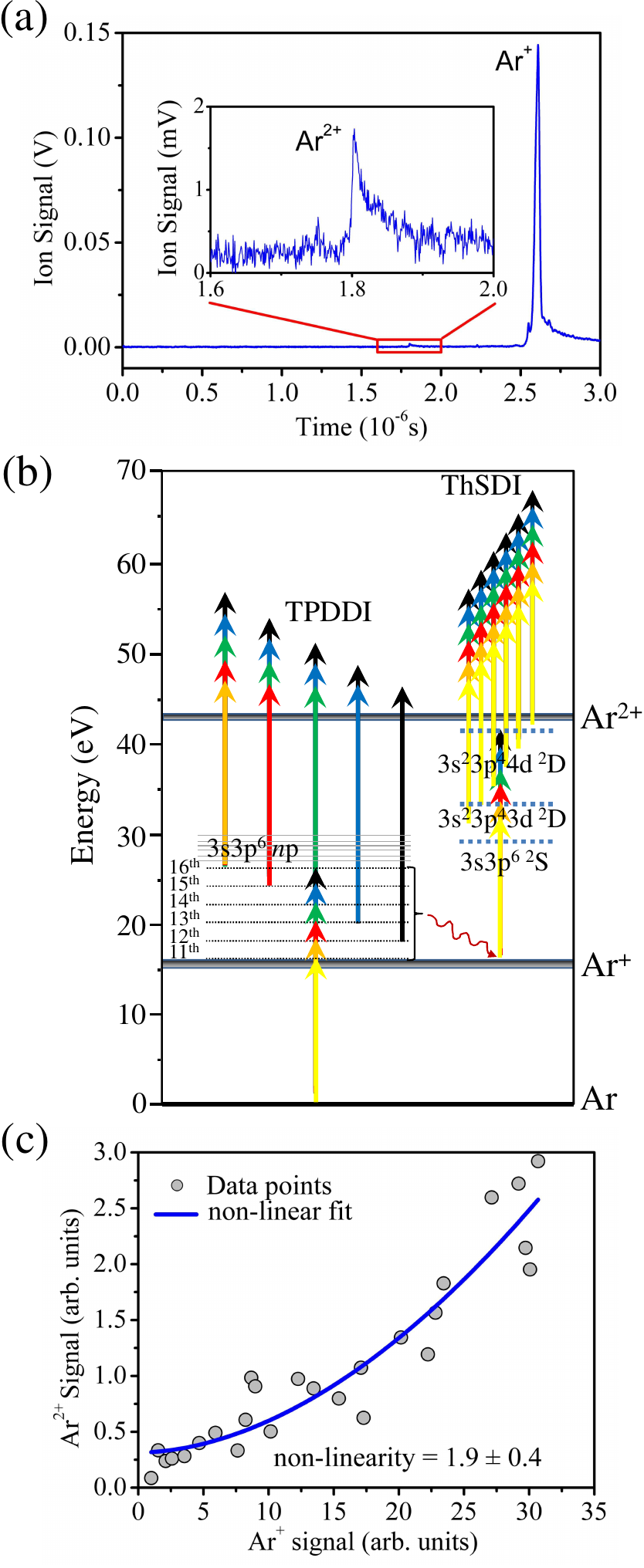} }
		\caption{ (a) TOF ion-mass spectrum recorded from long path harmonics ionizing TOF target (argon). $Ar^{2+}$ is shown in the inset. (b) Diagram of  $Ar^2+$ ion production. TPDDI two-photon-direct-double ionization and ThPSDI three-photon-sequential double ionization. In TPDDI the harmonics are far from the autoionizing (3s3p6np) states of Ar. In ThPSDI the harmonics 12th and 16th are near resonant with the 3s23p63d 2D and 3s23p64d 2D excited states of $Ar^{+}$. (c) $Ar^{2+}$ signal dependence upon XUV intensity. $Ar^{+}$ signal is produced via single-photon-ionization and thus is proportional to the XUV intensity. The change of the XUV intensity is achieved by changing the pressure of the harmonic generation medium. The offset of $Ar^{2+}$ signal observed around the zero values of $Ar^{+}$ signal is due to the offset of the ground signal of the oscilloscope.}
	\label{fig:Fig3}
\end{figure}
For the conditions used in the present experiment, the XUV intensity in the interaction region is smaller than the single photon ionization saturation intensity $I^{sat}_{XUV}\approx\frac{\hbar\time\omega_{XUV}}{\tau_{XUV}\time\sigma^{(1)}}\approx 7\cdot10^{12}\,W/cm^{2}$ ($\omega_{XUV}$ has been taken as the photon energy of the 14th harmonic). This is similar to the work reported in ref.\cite{benis1}, although the contribution of ThPSDI process cannot be excluded due to resonances with the excited state of $Ar^{+}$, the dominant excitation scheme is expected to be the one with the lowest order of non-linearity e.g. the TPDDI. The non-linearity of the double ionization process and the dominance of the TPDDI is confirmed by the quadratic non-linear dependence of the $Ar^{2+}$ signal upon the XUV intensity (Fig. 3c). In order to check if the TPDDI scheme can be used for the temporal characterization of the pulses via a $2-IVAC$ measurement we needed to evaluate further the contribution of each harmonic in the excitation scheme. The number of contributing channels scales from 5 for the $16^{th}$ harmonic down to zero for the $11^{th}$. Hence, the duration of the FTL XUV pulses in the train can be as short as $\sim 640\,as$. For this estimate, we assumed that the two-XUV-photon ionization cross section is constant along the bandwidth of the XUV radiation. The measured $2-IVAC$ trace shown in Fig. 4 results in an average duration that is 2.2 times longer than the FTL case i.e. $1.4\pm0.1\,fs$. This pulse broadening can be attributed to the chirp introduced by the metallic filter, the spectral phase distribution of the long path harmonics at the single-atom level \cite{hoffmann} and/or to spatio-temporal and shot-to-shot laser intensity variations in the harmonic generation medium \cite{kruse,gaarde,dubrouil}. 
\begin{figure}
	\centering
	\resizebox{0.85\columnwidth}{!}{%
	\includegraphics{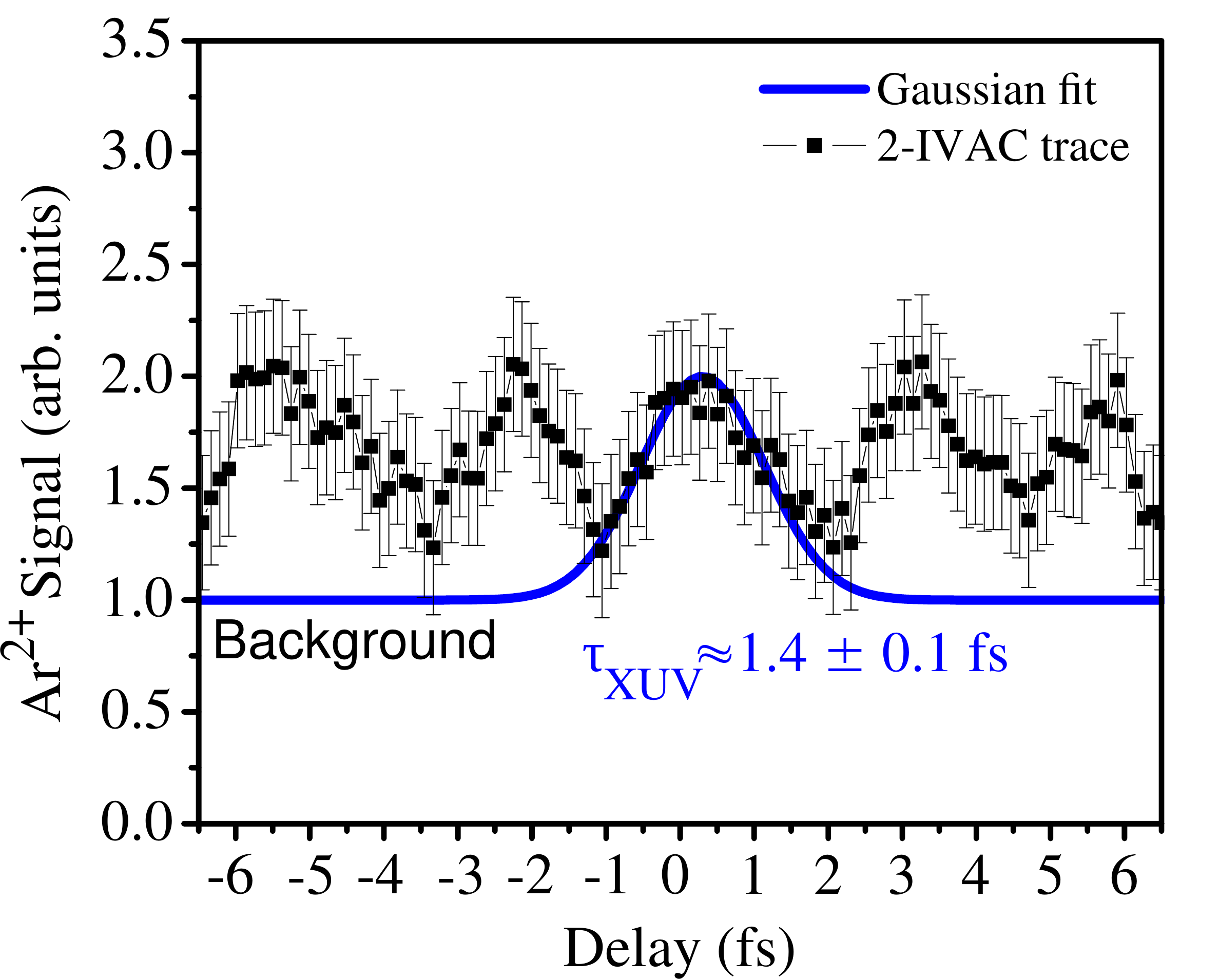} }
		\caption{2-IVAC trace (black dots) $Ar^{2+}$ signal produced by long path harmonics. Each delay corresponds to 100 shots accumulated. The XUV-XUV delay is feedback controlled via a piezoelectric translation stage of 1 nm step accuracy. The blue line resulted from fitting a sequence of Gaussian pulses to the raw data of the 2-IVAC trace. The pulses periodicity (repetition rate within the train) matches the HHG process i.e. 2.7 fs. The resulted average duration of the XUV pulses in the train is $1.4 \pm0.1\,fs$.  The baseline of the Gaussian fit has been placed at the signal value resulted when the XUV pulse replica are not spatio-temporally overlapped. The peak to background ratio of the trace is close to the value expected by the split mirror arrangement \cite{tzallas5}. }
	\label{fig:Fig4}
\end{figure}
Further work is needed to investigate the potential optimization of these pulses with the synthesized two-colour driving field parameter (inset of Fig. 2a) in line with study employed in \cite{brugnera,hoffmann}. These optical optimization of the wavefront could be appreciated via $2-IVAC$ measurements at different delays and intensity ratios between the two-colour of driving field.
\section{Conclusion}
\label{conclusion} 
Employing a two-colour driving field, we demonstrate the generation of intense long path high-order-harmonics. The high intensity of the XUV source has been confirmed by measuring the yield of $Ar^{2+}$ induced by a TPDDI process. Using this process we temporally characterized the long path harmonics by means of $2-IVAC$ technique.  We found that a superposition of six odd- and even-orders of long path harmonics results in a train of XUV pulses with average duration of $\sim 1.4\,fs$ which is essentially double the FTL duration. With these results we prove that long quantum path HHG can drive non-linear process and the corresponding EUV-XUV pulse duration is short enough to be exploited in a new route of EUV/XUV-pump-EUV/XUV-probe studies. The temporal characterisation over a range of $18-25 eV$ photon energy is important for studies on the absorption mechanisms in condensed matter such as photoionization of outer valence band electrons, inner electron ionization and Auger transitions. As shown in \cite{larsen}, due to the increased observation time-window provided, the long path high-harmonics are important for HHG spectroscopic studies of molecular dynamics, of relevance to decoherence studies \cite{vacher}.
 Our approach based on the combination of scaling invariant loose focusing geometry, two-colour field control, phase matching optimization and far field selection opens perspectives towards the generation of intense XUV pulses with controllable spectral phase distribution.  The generation of intense isolated XUV pulses by further combining with the polarization gating technique \cite{sola,sansone,tzallas4,mashiko} could be considered. 
\section{Acknowledgment}
\label{ack}
We acknowledge support of this work by the LASERLAB- EUROPE (EU’s Horizon 2020 Grant No. 871124), the IMPULSE project Grant No. 871161), the Hellenic Foundation for Research and Innovation (HFRI) and the General Secretariat for Research and Technology (GSRT) under grant agreements [GAICPEU (Grant No 645)] and NEA-APS  HFRI-FM17-3173. 
\\
We acknowledge support from EPSRC-CAF EP/J002348/1 CADAM, Royal Society RGS-R1-211053 and IES-R3-203022.
\section{Authors contribution}
\label{cont}
Conceptualization, D.C, P.T, JWG.T, JP.M and A.Z.; Methodology, P.T and A.Z; Software, PA.C, B.B and ES.S; Validation, PA.C, B.B and ES.S; Formal Analysis PA.C, B.B, ES.S and AZ; Investigation, PA.C, B.B, ES.S, D.C, P.T, JWG. T, JP.M and A.Z; Resources, D.C, P.T, JP.M and A.Z.; Data Curation, PA.C; Writing PA.C, P.T and A.Z; Writing, Review \& Editing, PA.C, B.B, ES.S, D.C, P.T, JWG. T, JP.M and A.Z; Visualization, PA.C, B.B, ES.S, D.C, P.T, JWG. T, JP.M and A.Z; Supervision, D.C, P.T, JP.M and A.Z.; Project Administration, A.Z; Funding Acquisition A.Z.

%\label{sec:1}
%and \cite{RefJ}
%\subsection{Subsection title}
%\label{sec:2}
%as required. Don't forget to give each section
%and subsection a unique label (see Sect.~\ref{sec:1}).
%
%\begin{figure}
% Use the relevant command for your figure-insertion program
% to insert the figure file.
% For example, with the option graphics use
%\resizebox{0.75\columnwidth}{!}{%
%  \includegraphics{fig1.eps} }
%\caption{Please write your figure caption here.}
%\label{fig:1}       % Give a unique label
%\end{figure}
%
% For tables use
%\begin{table}
%\caption{Please write your table caption here.}
%\label{tab:1}       % Give a unique label
% For LaTeX tables use
%\begin{tabular}{lll}
%\hline\noalign{\smallskip}
%first & second & third  \\
%\noalign{\smallskip}\hline\noalign{\smallskip}
%number & number & number \\
%number & number & number \\
%\noalign{\smallskip}\hline
%\end{tabular}
%\end{table}
%

\end{document}